\documentstyle[preprint,aps]{revtex}
\tightenlines
\begin{document}
\title{Neutrino Spectroscopy of the Early Phase of Nearby Supernovae}
\author{F. K. Sutaria$^*$, A. Ray$^{* \dagger}$}
\address{ $^*$Theoretical Astrophysics Group, Tata Institute of Fundamental
         Research, Bombay 400 005, India }
\address{ $^{\dagger}$Laboratory for High Energy Astrophysics, Goddard Space
Flight Center, Greenbelt, MD 20771} 
\address{ E-mail: fks@dashoo.tifr.res.in, akr@tifrvax.tifr.res.in}
\date{ Accepted for publication in Phys. Rev. Lett. on 11th Jul., 1997}
\maketitle
\begin{abstract}
\noindent
\label{sec:abstract}
Neutrinos emitted during stellar core collapse up to their 
trapping phase carry information about the stage from which the
Supernova explosion process initiates. The dominant $\nu_e$ emission mechanism 
is by electron capture on free protons and f-p shell nuclei and the spectrum of
these neutrinos is a function of the ambient physical conditions within the core
as well as the nuclear equation of state. The number
of collapse phase $\nu_e$ which can be detected by Super-Kamioka and Sudbury
Neutrino Observatory from a Supernova within 1 kpc, and their generic
energy spectra are given.
\noindent
\vskip 0.01 true in
Ms number ~~~~~~~~~PACS number: 97.60.Bw, 95.55.Vj, 26.50.+x, 95.85.Ry \\
\end{abstract}
\newpage
\section{Introduction}
\label{sec: Intro}
Although several dozen Supernovae are routinely detected in external
galaxies every year in the optical channel, 
the small interaction cross-section of neutrinos with matter together
with the enormous dilution of their flux due to large distance of external
galaxies has prevented the detection of Supernovae in the neutrino channel
with the exception of SN1987A. However, since large experiments with 
the main purpose of detecting Solar neutrinos are either already on
line (such as Super-Kamioka (SK)) or expected to become operational soon
(e.g. Sudbury Neutrino Observatory (SNO)) it is pertinent to consider 
what such experiments can reveal about late stages of stellar evolution,
such as the gravitational collapse of a massive star leading to a
Supernova - should such an event occur in our galaxy close to the Earth
$^{\cite{Burr},\cite{RSS97}}$.

The 19 neutrinos detected from SN1987A were most likely to have been
emitted during the post-bounce phase as their total fluence
during the proto neutron star cooling phase 
(at $\simeq 10^{58}$)
is much larger than that during the collapse phase($\simeq 10^{56}$). 
Much of the information pertaining to the conditions in which the
neutrinos are originally produced, such as the nuclear
and thermodynamic properties of the core of the Supernova are altered
because these neutrinos undergo {\it inelastic}
scattering with the overlying stellar matter
in the post neutrino trapping phase.
Neutrinos which are emitted through electron captures
on the nuclei present in the pre-supernova and collapsing core before it
reaches neutrino trapping density ($\simeq 10^{12}$gm/cm$^3$)$^{\cite{Arn77},
\cite{Bru}}$,
however, stream freely through the stellar
matter without any further interactions. These pre-trapping neutrinos
carry with them information on both the physical conditions within the
core, as well as it's nuclear configuration e.g. the ratio of
the number density of free protons to that of 
heavy nuclei. 
The last quantity can be 
dependent on the nuclear equation of state 
relevant to this region of collapse. 
Since neutrinos act as probes of
the dynamic, thermodynamic and nuclear properties of the pre-supernova and
collapsing core, their detection and measurement of
energy spectrum can have 
significant implications.
The time evolution 
of the detected spectrum could also reveal the dynamical timescale -- a
clue to the average density and mass of the stellar core 
which may have implications for black
hole vs neutron star formation. 
The collapse phase emission is in the $\nu_e$ channel. Since 
the $\nu_{\mu}$ and $\nu_{\tau}$ flavors can be detected only by 
neutral current (n.c.) reaction on $d$ nuclei in SNO 
or by $\nu-e^-$ scattering, any excess of neutrinos
detected in the n.c. channel (over the value measured from
the charged current (c.c) channel) in SNO  would 
be indicative of neutrino oscillation.
The extent of the reduction of lepton fraction during stellar
collapse has important implications for shock formation stages
and the overall dynamics - 
even in the delayed explosion stage, since it
determines, through the original
energy of the bounce shock, and the entropy profile
in the outer core, the position of the stalled shock
\cite{Jan97}.

\section {Number of neutrinos emitted and detected}
\label{sec: Number} 
Since electron captures take place throughout the mass of the 
iron core, 
the fluence of $\nu_e$ received per unit area at a distance D
up to a certain
stage characterized by the mass-weighted electron fraction (Y$_{ef}$) in 
the core, is given by 
$F_{\nu_e}= M_{C} \Delta Y_e / (4 \pi D^2 m_B)$ ,
where $M_{C}$ is the mass of the collapsing ``Fe" core, 
m$_B$ the baryonic mass,
and $\Delta Y_e = Y_{e_f} - Y_{e_i} $ is the change in the lepton fraction
from the ``start" of collapse (corresponding to the end of Kelvin-Helmholtz 
contraction following Si exhaustion) 
to the point where $\nu_e$ begin to be trapped and
scattered by overlying matter in the collapsing core.
Stellar core density  
of $\approx 2 \times 10^{11} \rm g/cm^3$ is still transparent
for 10 to 20 MeV neutrinos.
The spectrally integrated fluence of $\nu_e$
at a distance of 1 kpc, 
as $Y_e$ changes from 0.42 to 0.39 in a 1.4 M$_{\odot}$ stellar core
(of a 15 M$_{\odot}$ star)
is: $F_{\nu} = 4.2 \times 10^{11} \rm cm^{-2}$.
The energy of the infall 
neutrino burst up to this stage is:
$ E_{\nu_e} = 7.2 \times 10^{50} \rm erg $.

The flux, direction and the spectra of the 
neutrinos can be measured by
the charge current dissociation of the deuterium nucleus 
($\nu_e(d,pp)e^-$) in 
the Sudbury Neutrino Observatory (SNO) $^{\cite {SNO}}$ 
with a fiducial mass of 1 Kt
of high purity $\rm{ D_2 O}$.
It is also possible to detect $\nu_e$  and obtain spectral information 
by means of the neutrino-electron scattering reaction 
($\nu_e +e^- \to \nu_e+ e^-$) in SNO as well
as in the light water detector Super Kamioka. 
Apart from the reactions mentioned above which can
perform spectroscopy (i.e. measure the incoming neutrino energy),
the neutral current dissociation of $d$ 
by the reaction $\nu_e(d,pn)\nu_e$ can obtain the 
the total neutrino flux (of all flavors), 
but since it does not give any neutrino energy information,
it is equivalent to neutrino "photometry". 

The number of $\nu_e$ events which can be expected to be detected in the SNO  
detector through neutrino-induced 
c.c.  reaction on the target $d$ nuclei  is given by:
 $n_{\nu_e} = F_{\nu} \sigma_{cc}(\epsilon_{\nu_e}) N_{d}$
where $N_d$ $(=6.02 \times 10^{31})$ is the total number of target nuclei 
present in the $1$Kt detector.
The charge current  and neutral current cross-sections 
($\sigma_{cc}(\epsilon_{\nu_e})$ and $\sigma_{nc}(\epsilon_{\nu_x})$ 
respectively),  
have been computed 
for the $\nu$-$d$ process
by Bahcall et al. \cite{Bah88}
and accurate fits to these cross-sections between 5 to 40 MeV
are given as$^{\cite{Burr}}$:
$ \sigma_i = \alpha_i ( \epsilon_{\nu} - \epsilon_{th, i})^{2.3} $
where $i=cc$ and $nc$, $\alpha_{cc} = 1.7 \times 10^{-44}$ cm$^2$,
$\alpha_{nc} =0.85 \times 10^{-44}$ cm$^2$, $\epsilon_{th,cc} =
2.2$ MeV and $\epsilon_{th,nc} = 1.44$ MeV.

For the H$_2$O based Cerenkov detector
(Super-Kamioka) the ($\nu_e, e^-$) scattering events would be 
the main source of $\nu_e$ spectral information since the corresponding
energy thresholds for charge current and neutral current interactions for
ordinary water  are much higher.
(During the collapse stage, the neutrino flux is almost entirely
in neutrinos of the electron type; 
anti-neutrinos of various kinds, as well
as neutrinos of the mu or tau type are generated in copious numbers
only in the hot post core bounce phase).
The relevant ($\nu_e, e^-$) scattering cross-section is$^{ \cite{Seh74}}$ :
$ \sigma_e = (1/2)(4G^2 m_e^2 \hbar^2 / \pi c^2) (7/12)
(\epsilon_{\nu} / m_e c^2) $

The number of detections by Super-Kamioka (mass 32 kt) and
SNO  for a supernova explosion 1 kpc away, for several possible
scenarios of stellar core collapse are reported
in Table 1. 
The 15 M$_{\odot}$ star collapse is initiated from thermodynamic
conditions as in \cite{Full82} (Y$_{ei}$ = 0.420, $S_i/k_B$ = 1.00,
T$_i$ = 0.7178, $\rho_{10}$ = 0.37), while the 25 M$_{\odot}$ star's
single zone initial conditions are similarly derived from the data 
reported in \cite{WWF} and an expression
for the core averaged entropy (Y$_{ei}$ = 0.423, $S_i/k_B$ = 1.14,
T$_i$ = 0.6756, $\rho_{10}$ = 0.15).

\section{Incident neutrino Spectrum:}
\label{sec:Spectra}
To estimate the number of neutrinos incident upon an underground detector,
at a distance 
$D$ kpc from the SN, we use the stellar core collapse code developed in 
ref.\cite{RCK}. This  code uses the "single-zone" method
outlined in \cite{BBAL} (hereafter BBAL) 
exploiting the homology of the core structure during collapse but
used the electron capture rates for both heavy nuclei and free protons
with temperature dependent phase space factors and neutron
shell-blocking in nuclei included where relevant, 
an analytic equation
of state for hot dense matter as in {\cite{Full82}
and a treatment of the neutrino trapping and leakage at densities above a few
times 10$^{11} \rm g/cm^3$.
The core of a massive star collapses under its own gravity when
the pressure support from degenerate electrons is reduced through
the capture of electrons 
in the stellar material. The electron capture on neutron rich 
heavy nuclei 
initially
proceeds primarily through the allowed type
($\Delta l$ = 0) Gamow Teller transitions. 
As core density exceeds $\simeq 10^{11} \rm gm/cm^3$, the nuclei 
become become more and more massive and too neutron rich to allow 
e$^-$-capture to take place through allowed Gamow-Teller transitions
from the ground state.
This is because the allowed states for $p$ to $n$ transition within
the nucleus are already filled by the neutrons (neutrons shell blocked) 
and the transition strength for
typical captures like $^{56}$Fe $\to$ $^{56}$Mn used earlier (as in
\cite{BBAL}) is
no longer representative of typical nuclear e$^-$-capture rates. 
It was estimated \cite{Full82} that the effective transition strength 
for electron capture under
these circumstances due to first forbidden transitions and due to 
thermally unblocked allowed transitions. 
However it was shown$^{\cite{KR83},\cite{Zar}}$ that the dominant
unique first forbidden transition strength was actually negligible
compared to the thermally unblocked strength under the typical
core collapse conditions. Therefore, after neutron-shell blocking sets in,
(when (A,Z) $>$ $^{74}$Ge) 
the sum rule for the Gamow Teller transition operator $|M_{GT}|^2$
decreases from a typical value of 2.5 $^{\cite{BBAL}, \cite{Full82}}$
to about 0.1.
The e$^-$-capture rate on a single nucleus 
$X(A,Z)$
in the initial state $i$ to the 
final state j
is given by:

$$ \lambda_{ij} = \ln 2 {f_i(T,\mu_e, Q_{ij}) \over ft_{ij}}$$
where $ft_{ij}$  is related to $|M_{GT}|^2$  
by $ft_{ij}= 3.596 - \log |M_{GT}|^2$ for 
allowed Gamow-Teller type transitions (for free protons,
log ft$_{f.p.}$ = 3.035).
The factor $f_i(T,\mu_e, Q_{ij})$ is the phase space factor for
the allowed transition, which is a function of the 
ambient temperature $T$, the Fermi-energy of the electron $\mu_e$ and 
the Q-value for the reaction $Q_{ij}$.  
>From energy conservation, the neutrino energy is
$E_{\nu_e} = E_e - Q_{ij}$. 

In the calculations performed here, 
we follow the self-similar
density evolution of a single zone which is represents the
properties of the polytropic distribution of  matter in the core and,
in which the density, entropy per nucleon, and electron fraction
are allowed to evolve self-consistently with the electron
capture physics and the consequent changes in nuclear and thermodynamic
variables. 
During the collapse, 
a change in entropy controls 
the fraction of the dripped protons with respect to that of the heavy
nuclei, and as the spectrum of neutrinos generated by electron
capture on protons are different from captures on heavy nuclei,
this influences the overall neutrino spectrum received on earth.
The received neutrino spectrum may depend not only upon
the initial conditions from which the collapse started, 
but also 
on the details of the electron capture properties of the stellar
matter. Properties of nuclei at finite temperatures and density
during this phase of the collapse, where shell and pairing corrections
are relevant, are being computed \cite{NIC96}. 

The rate of generation of neutrinos per nucleon in a given energy band
$E_{\nu}$ to $E_{\nu} + d E_{\nu}$ 
taking account of relative abundance of free protons and nuclei,
is given by:
$$ dY_{\nu}(E_{\nu}) = d\lambda_{fp}(E_{\nu}) X_p  + 
                       d\lambda_{H}(E_{\nu}) (1 - X_n - X_p)/A $$ 
\noindent where A represents the Atomic weight of the ensemble of  nuclei
present in the core, taken here to be represented by a single
"mean" nucleus as in earlier works \cite{BBAL}. The differential neutrino
production rates for free protons and heavy nuclei are:
$$ d\lambda_{fp, H} = { \log 2  \over (ft)_{fp,H} } {<G> \over (m_ec^2)^5}
{ E_{\nu}^2 (E_{\nu} + Q_{fp,H}) \sqrt{ (E_{\nu} + Q_{fp,H})^2 - (m_e
c^2)^2) } \over (1 + \exp( E_{\nu} + Q_{fp,H} - \mu_e)) } d E_{\nu} $$ 
where the Coulomb correction factor $<G>$ has been taken
as $\approx 2$  for heavy nuclei and 1 for free protons. 
The Q-value, assuming that the strength is concentrated in a single
state, is given as:
 $Q = (\hat \mu + 1.297 + E_{GT})$ 
where $\hat \mu$ ($=\mu_n - \mu_p$) is the difference in the neutron and 
proton chemical potentials 
when free nucleons coexist with a 
distribution of neutron rich nuclei in nuclear statistical equilibrium, 
and $E_{GT}$ is the energy of the Gamow-Teller Resonance centroid.
The centroids in fp-shell nuclei, found from experimental data
from (n,p) reactions have been used for characterizing GT transitions in
these nuclei $^{\cite{SR95}}$ and are close to the value
(3 MeV) used here.
The Fermi-energy 
$\mu_e = 11.1 (\rho_{10} Y_e)^{1/3} \rm MeV $. The
difference in chemical potentials $\hat \mu$,
and the relative fraction of free protons 
are obtained from a low density analytic equation of state
similar to that in \cite{BBAL} with modifications noted in \cite{Full82}.

We present a typical (``snapshot")
spectrum of neutrinos from a 15 M$_{\odot}$ star's core collapse
in Fig 1 within a narrow
range of stellar core density around $10^{11} \rm g/cm^3$. 
Note the two separate peaks due to neutrinos from
capture on free protons and heavy nuclei
with clearly non-thermal spectrum.
On the other hand, in Fig 2(a)
we present a cumulative spectrum of neutrinos from the same
star emitted till the
stellar core density reaches $2.4 \times 10^{11} \rm g/cm^3$
(typically
120 milliseconds after the onset of collapse, and about 10 milliseconds
before core bounce). The broader distribution of the neutrinos from
captures on free protons reflects the full range of electron Fermi
energies extant due to the variation of stellar core density
as the collapse proceeds. 
Allowed capture rate on heavy nuclei is strongly attenuated, once
the stellar core reaches a density of approximately 10$^{11} \rm 
g/cm^3$
at which point the nuclei enter the neutron shell blocked phase.
The range of density for which heavy nuclear capture effectively 
persists is therefore somewhat more restricted.
As a result the cumulative spectrum shows a substantial fraction
of high energy neutrinos.
As  the expected number of neutrinos detected depends strongly
on the overall fraction of the high energy neutrinos from
proton capture, some conditions of stellar collapse may be easier
to detect than others. Figures 2(b) and 2(c) gives the
expected number of detections in the SNO and Super-Kamioka
detectors, after folding the incident spectrum through
the relevant interaction cross-sections assuming  
uniform and full detection
efficiency.
The spectra shown in Figs 1 and 2 are calculated with an effective
sum-rule of the matrix element $|M_{GT}|^2 = 1.2$ and 
reduced only when (A,Z) =$^{74}$Ge is reached.
The former value for unblocked nuclei 
is characteristic of shell model calculations
of $\beta$-strength function in iron isotopes \cite{Bloom}.
We have also calculated
spectra with $|M_{GT}|^2$ =2.5 in the unblocked era and found
the cumulative spectra qualitatively similar, although somewhat
deficient in higher energy neutrinos (results summarized
in Table 1). 
The  set of mean-nucleus transition
matrix elements used for generating Figs. 2(a), (b) and (c) 
may be more realistic, because the A distribution of nuclei in collapsing
stellar core is rather broad, and even at the onset of
stellar collapse, the mean nucleus is fairly close to the
neutron shell-blocked configuration. Therefore a substantial
fraction of the nuclei present in the core 
may already have low capture transition strengths at the beginning of collapse. 
This fraction becomes larger
after the mean nucleus enters the shell-blocked phase. 
Although the spectra have been obtained in the ``mean nucleus"
approximation while in reality the stellar core consists of an
ensemble of nuclei, only a few nuclei with the largest abundance and the
smallest effective Q-value (allowed) are expected to dominate the heavy
capture spectrum (see e.g. \cite{Coop}).
Results of another example (core collapse of a 25 M$_{\odot}$ star)
are summarized in Table 1. This case had a higher core averaged entropy
than the 15 M$_{\odot}$ case, and as a result has a larger
fraction of free protons in the core. Consequently the cumulative
spectrum 
is richer in high energy neutrinos.

To compare the numbers expected from the pre-trapping phase with the
post core bounce phase, note that the number of neutrinos of
all flavors in the post bounce phase from a SN 1 kpc away is
expected to be 592,000 in Super Kamioka (550,400 from $\bar \nu_e(p,n)
e^+$ and 41,600 from all types of neutrino electron scattering);
similarly, 51,500 (neutrinos
of all flavors and their anti particles) are expected
in SNO (see eqs. 7.A.3,
7.A.4 and 7.A.7 of \cite{Burr}).

Neutrino spectroscopy of the final state of a star
would be possible
provided that the event occurs at a relatively close distance. 
Although, a priori, these may be rare events, there have been a
number of historical Supernovae, as well as
detected radio pulsars within a distance of about 2 kpc.
There are a number of star forming regions nearby (such as
the Orion complex - about 440 pc away), 
which are sites of potential supernova progenitor.
A detection by underground neutrino experiments
could constrain features of theoretical
calculations of both collapse and explosion era of type II
Supernovae as well as shed light on the 
characteristics of the stellar core.

This work was partially supported through the 8th Five Year Plan Project
8P-45 at Tata Institute. A.R. was supported by a National Research
Council Senior Research Associateship at NASA/Goddard Space Flight
Center.

\begin{table*}
\begin{center}
\caption{Pre-trapping
neutrino detections in SNO and Super Kamioka with hardness ratios
up to $\rho_{10}$ = 24.16 for indicated heavy nuclear e-capture
matrix elements for 15 M$_{\odot}$ Fuller (1982) and 25 M$_{\odot}$
WWF presupernova stars.}
\vskip 1 true cm
\begin{tabular}{c|c|c|cc|cc|cc}
\tableline
Star Mass  &$|M_{GT}|^2  $ &$t_{collapse}   $ &
\multicolumn{2}{c|} {Pre-trapping Variables}&
\multicolumn{2}{c|} {No. Detected} &
\multicolumn{2}{c|}{Hardness Ratio$^{\dagger}$} \\
& & (ms) & $Y_{ef}$& $S_f/k_B$ & 
SNO & S-K & SNO & S-K \\ 
15 M$_{\odot}$ & 1.2/0.1 &120 & 0.3909   & 1.0021   & 82     &394       &0.2786 
&0.8540  \\
               & 2.5/0.1 &120 & 0.3896   & 1.0085   & 66     &344       &0.2876 
& 0.9537  \\
25 M$_{\odot}$ & 1.2/0.1 &190 & 0.3828   & 1.1080   &120     &566       &0.2878
&0.8319  \\
               & 2.5/0.1 &190 & 0.3813   & 1.1204   & 99     &499       &0.2916 
&0.9190 \\
\end{tabular}
\end{center}
$^{\dagger}$ The hardness ratio denotes the number
of neutrino events in the 5 MeV $\leq E_{\nu_e} \leq 12$ MeV
and 12 MeV $\leq E_{\nu_e} \leq$ 25 MeV bands.
\end {table*}

\begin{figure}
\caption{``Snapshot" 
neutrino fluence (MeV$^{-1}$ per baryon in a narrow density interval 
$\Delta\rho_{10} = 0.0002$
around  $\rho_{10}=9.8668 \rm gm cm^{-3}$ for a 15 M$_{\odot}$ star at
D= 1 kpc and $|M_{GT}|^2$ = 1.2 and later 0.1.}
\end{figure}
\begin{figure}
\caption{(a) Cumulative neutrino fluence up to  $\rho_{10}=24.16
\rm gm cm^{-3}$, with M = 15 M$_{\odot}$, 
D = 1 kpc and $|M_{GT}|^2$ = 1.2 and later 0.1.
(b) The spectrum in (a) folded with the detection cross-section for 
c.c. reaction $\nu_e(d,pp)e^-$ in SNO.
(c) The spectrum in (a) folded with the detection cross-section 
for $\nu_e -e^-$ scattering in Super Kamioka.}
\end{figure}

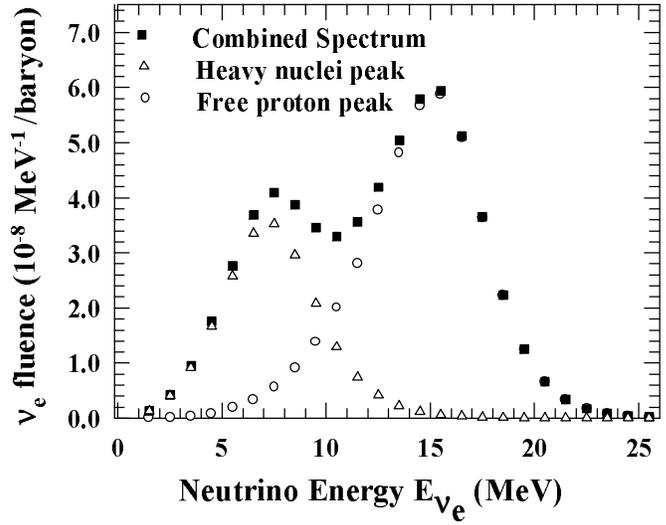
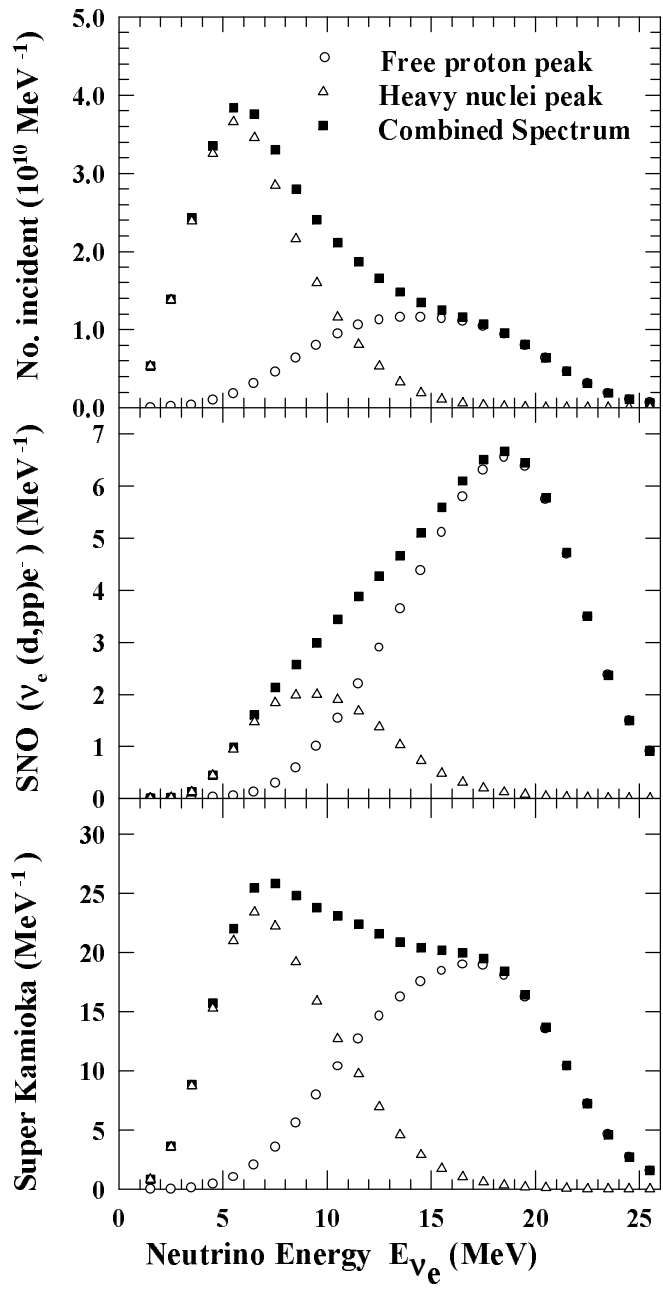
\end{document}